\documentclass[sn-mathphys,Numbered]{sn-jnl}

\usepackage{graphicx}%
\usepackage{multirow}%
\usepackage{amsmath,amssymb,amsfonts}%
\usepackage{amsthm}%
\usepackage{mathrsfs}%
\usepackage[title]{appendix}%
\usepackage{xcolor}%
\usepackage{textcomp}%
\usepackage{manyfoot}%
\usepackage{booktabs}%
\usepackage{algorithm}%
\usepackage{algorithmicx}%
\usepackage{algpseudocode}%
\usepackage{listings}%
\usepackage{makecell}
\usepackage{bm}
\usepackage{soul}

\newcommand{\be}{\begin{equation}}
\newcommand{\ee}{\end{equation}}
\newcommand{\bs}{\begin{split}} 
\newcommand{\bea}{\begin{eqnarray}}
\newcommand{\eea}{\end{eqnarray}}

\begin{document}

\title[Article Title]{On the weak and strong field effects in antiscalar background}

\author[1]{\fnm{Eduard} \sur{Mychelkin}}\email{mychelkin@fai.kz}

\author*[1]{\fnm{Maxim} \sur{Makukov}}\email{makukov@fai.kz}

\author[1,2]{\fnm{Gulnara} \sur{Suliyeva}}\email{suliyeva@fai.kz}

\author[1,2,3]{\fnm{Nosratollah}
\sur{Jafari}}\email{nosrat.jafari@fai.kz}

\affil*[1]{\orgname{Fesenkov Astrophysical Institute}, \orgaddress{\street{Observatory 23}, \city{Almaty}, \postcode{050020}, \country{Kazakhstan}}}

\affil[2]{\orgname{Al-Farabi Kazakh National University}, \orgaddress{\street{Al-Farabi av. 71}, \city{Almaty}, \postcode{050040}, \country{Kazakhstan}}}

\affil[3]{\orgname{Center for Theoretical Physics, Khazar University}, \orgaddress{\street{41 Mehseti Street}, \city{Baku}, \postcode{AZ1096}, \country{Azerbaijan}}}


\abstract{

The triumph of general relativity under the banner ``gravity is
geometry'' began with confirming the crucial effects within the Solar
system and proceeded recently to the strong-field shadow effect for
the compact object in the center of the Milky Way. Here, we examine
some of those phenomena for the Einstein-scalar equations in the
antiscalar regime to reveal the difference from vacuum both in weak
and strong fields. As a result, we find that for week-field perihelion
shift the difference between vacuum and antiscalar cases proves to be
observationally imperceptible in practice, even for S-cluster stars
with high eccentricities, and even if accumulated over a century. In
strong-field case, we reconsider the shadow effect (this time without
involving complex-valued scalar field) as the most perspective from an
observational viewpoint. Even though the resulting difference is quite
appreciable (about 5\%), no conclusion can be made until the mass of
the central object is known with the accuracy an order of magnitude
higher than the currently available.

}





\maketitle

\section{Introduction}\label{sec1}

The mainstream in the study of scalar field effects for a long time has been represented by the Janis-Newman-Winicour (JNW) family of static solutions of the Einstein-scalar equations (ESE) with the usual sign of minimal scalar field EMT \cite{Fisher1948,JNW}.  However, the solutions of this family prove to be unstable with respect to collapse (see, e.g., \cite{Faraoni21}), and so, their physical meaning is questionable.

The  exponential Papapetrou solution of the ESE is obtained when the background scalar field is taken in antiscalar mode, i.e. with negative sign of its energy-momentum tensor (EMT) \cite{papa54,Yilmaz1958}. The motivation for considering alternative antiscalar model (with the sign flip of the scalar EMT) lies, apart from its stability, in the fact that it conforms to observational data not worse than the standard vacuum solution, leading, at the same time, to considerable simplification of local properties of the field sources (as compared to the vacuum case, see \cite{mm18}).

Historically, it is interesting that even though Papapetrou was the first who derived the metric \cite{papa54}, it was only Yilmaz \cite{Yilmaz1958} who first indicated, in a footnote, that exponential metric follows from the opposite sign of the minimal scalar field EMT. Later, by analogy with anti-de Sitter vacuum (`negative $\Lambda$-term'), this metric was dubbed `antiscalar'.

The term `antiscalar field' should not be mistaken for `phantom field'. The latter stands for traditional cosmological phantom  scalar field with (problematic) equation of state described by the state parameter $w < -1$. As for the antiscalar regime, it does not imply any new exotic EMT with a new equation of state; it only requires the opposite sign for the \emph{standard} minimal scalar EMT entering the ESE (justified, again, by excellent agreement with observations). The energy conditions for the EMT with this metric have been considered (with extension to rotation) in \cite{mm23} on the basis of the Hawking-Ellis energy criterion of type I. 

Along with the well-known solutions of the ESE described in \cite{papa54,Yilmaz1958} (together with corresponding Klein-Gordon equation) there are two operational methods for derivation of the antiscalar exponential metric (together with corresponding potential function) \cite{mm18,mm20}. The first one is based on employing purely imaginary scalar field instead of the original real-valued one in the JNW approach, leading to corresponding antiscalar results in all analytical consequences (for details see \cite{mm18}). The second method is based on adopting the scalar charge to be equal to the mass of the source in the scalar solutions, thereby operating with only real scalar field, which is automatically transferred into antiscalar regime (for details see \cite{mm20}).
 
Note an important consequence of the latter method: since the source of the field  (scalar charge) reduces exactly to the central mass value, this scalar charge cannot be treated as a small perturbation. Moreover, such approach assumes possibility for the description of gravity as a phenomenon \emph{de facto} conditioned by scalar field. Thereby, the dilemma ``vacuum vs. scalar background'' might have a fundamental meaning for the interpretation of general relativity .

Recently, it had been shown that the exponential metric represents a traversable wormhole, with ``all of the interesting and potentially problematic'' consequences \cite{visser}. This should not make a difference as per the observational effects calculated from the metric, and exactly observations should decide which model (vacuum or antiscalar) is best matched. While they both excellently agree with one-shot weak-field observational data, the hope is for future data on strong-field effects, as well as weak-field effects accumulating in time.

Hence, here we analyze, for scalar and vacuum background, the observable phenomena such as perihelion precession effects for (1) planet Mercury and (2) four S-cluster stars -- S2, S38, S55, S62, and also (3) reconsider the shadow effect for compact objects in strong field case.

\section{Equations of motion}\label{sec2}

To study the motion of a particle in both scalar background and vacuum
we use the same general isotropic coordinates, 
   \be  \label{is}  ds^2= B(r) dt^2 - D(r) \left( dr^2 +  r^2 d\theta^2
       +  r^2 \sin^2 \theta d\phi^2\right) .            \ee
To distinguish scalar background from vacuum we write, where
needed, index `P' for the Papapetrou anti-scalar and index `S' for the Schwarzschild vacuum cases: 
       \be \label{P}   B_P(r)= e^{-2\varphi}= e^{-2 M/r}, \qquad D_P(r)= e^{2\varphi}= e^{2 M/r} \ee
      and
       \be \label{S}   B_S(r) = \left( \frac{ 1-\frac{M}{2 r}}{1+\frac{M}{2 r}}
       \right)^2, \qquad D_S(r) = \left(1+\frac{M}{2 r} \right)^4. \ee    
Equations of motion for a test particle can be derived from the Lagrangian
       \be \label{Lagrangian}
       \mathcal{L} = \frac{1}{2}  g_{\alpha \beta}(x) \dot{x}^{\alpha}\dot{x}^{\beta} = \frac{1}{2}\left[B(r) \dot{t}^2 -D(r)\left( \dot{r}^2 +  r^2 \dot{\theta}^2
       +  r^2 \sin^2 \theta \dot{\phi}^2\right) \right]
       \ee
       using the Euler-Lagrange equations, $\frac{d}{d\tau}\left(\frac{\partial \mathcal{L}}{\partial \dot{x}^{\alpha}}\right) - \frac{\partial \mathcal{L}}{\partial x^{\alpha}} = 0$. Then, one obtains:
    \be \label{eqt}    \ddot{t} + \frac{B'}{B}\dot{r}\dot{t} = 0, \ee    
    \be \label{eqr}    \ddot{r} + \frac{B'}{2 D}  \dot{t}^2 + \frac{D'}{2 D} \left( \dot{r}^2  -  r^2 \dot{\phi}^2 \right) -
    r \dot{\phi}^2  =0 , \ee\
    \be \label{eqphi}    r^2 D \left[\ddot{\phi} + \left(\frac{2}{r} + \frac{D'}{D}\right)\dot{r} \dot{\phi} \right]=0. \ee
Integration of \eqref{eqt} and \eqref{eqphi} leads to:
    \be    \label{8}  B(r) \frac{dt}{d \tau}= B \dot{t} =
       \text{const},  \quad D(r) r^2 \frac{d\phi}{d \tau}=J,\ee
where $J$ is a constant,  $\tau$  is the
proper time, overdot denotes derivative with respect to $\tau$
and prime -- with respect to $r$. From those it follows:
 \be  \label{16}  \frac{1}{D r^4} \Big(  \frac{dr}{d\phi} \Big)^2 +  \frac{1}{ D r^2 }  -   \frac{1}{ J^2 B} = -\frac{E}{J^2},   \ee
   \be        r^2 \frac{d\phi}{dt}=\frac{J B }{D}, \quad ds^2= E B^2 dt^2,   \label{eb2} \ee
   where $E$ is a constant proportional to the specific energy of a test body.
The solution of (\ref{16}) is a direct integration:
   \be \label{19}   \phi=   \pm \int  \frac{ dr }{r^2  \Big[   D/(J^2 B) - ED/J^2- 1/r^2    \Big]^{1/2}  }\, ,    \ee
with $B=B(r)$ and $D=D(r)$. We adopt the clockwise motion to
be in the direction opposite to standard angle $\phi$ increasing
counter-clockwise, so we choose the `minus' sign.

\section{General results for the precession of perihelia}\label{sec3}

From Eq.~\eqref{16}, under the
condition $dr/d\phi = 0$ defining perihelion $r_-$ and aphelion $r_+$,
i.e.
\begin{equation}
\frac{1}{ D^\pm r^2_\pm }  -   \frac{1}{ J^2 B^\pm} =
-\frac{E}{J^2},  \qquad B^\pm \equiv B(r_\pm), \quad  D^\pm \equiv D(r_\pm),
\end{equation}
in arbitrary isotropic coordinates the constants of motion $E$ and $J$
are found as: 
\be \label{EJ}     E=\frac{     \frac{D^{+} r_{+}^2  }{ B^{+} }   -
	\frac{D^{-} r_{-}^2  }{ B^{-} }    }{  D^{+}  r_{+}^2 -  D^{-}
r_{-}^2  },  \qquad   J^2 =  \frac{  \frac{1}{B^{+}}- \frac{1}{B^{-}}  }{
	\frac{1}{ D^{+} r_{+}^2  }   -  \frac{1}{ D^{-} r_{-}^2  } }.   
\ee
We can also express $r_{-} $ and $r_{+}$ through the
parameters of an elliptical orbit, the eccentricity $e$ and semi-major
axis $a$: $r_{\pm} = (1\pm e)a.$

With indicated constants fixed in corresponding isotropic
 coordinates, the angle swept from $ r_{-} $ by the position vector
 $r$ follows from \eqref{19} exactly as
  \be   \phi(r)- \phi (r_{-})  = - \int_{r_{-}}^{r}  
  \Big[ \frac{D(r)}{ J^2 B(r)  } -  \frac{E D(r)}{ J^2  }  - \frac{1}{
  r^2   }     \Big]^{-\frac{1}{2}}    \frac{dr}{ r^2  } .
\label{diffpsi}
\ee
When the expression under the square root is calculated to the second order in
$M/r$, it might be represented in the following form (cf. \cite{Weinberg1972}):
\begin{equation}
     \frac{D(r)}{ J^2 B(r)  } -  \frac{E D(r)}{ J^2  }  - \frac{1}{
         r^2   }  = C \left( \frac{1}{r_{-}} - \frac{1}{r} \right)
         \left( \frac{1}{r} - \frac{1}{r_{+}} \right).
\end{equation}
For this expression to be valid for all $r$, the constant $C$ follows
from the limit $r \to \infty$:
\begin{equation}
    C = \frac{r_{-}r_{+}}{J^2} \left( E-1\right),
     \label{C}
\end{equation}
where $C$, $J$, and $E$ should be taken separately for scalar background
($C_P$, $J_P$, $E_P$) and for vacuum ($C_S$, $J_S$, $E_S$). As a
result, \eqref{diffpsi} assumes the form:
\begin{equation}
    \phi(r) - \phi(r_{-}) = \frac{- J}{\sqrt{r_{-}r_{+} \left(
    E-1\right)}} 
    \int_{r_{-}}^{r}  
      \left[  \left( \frac{1}{r_{-}} - \frac{1}{r} \right)
         \left( \frac{1}{r} - \frac{1}{r_{+}}
     \right)\right]^{-\frac{1}{2}}
         \frac{dr}{ r^2  }.
\label{diffpsi2}
\end{equation}
The standard change of variables \cite{Weinberg1972}
\begin{equation}
    u=  \frac{1}{r} = \frac{1}{L} + \frac{1}{N} \sin\psi,
  \qquad
  du = -\frac{dr}{r^{2}}=\frac{1}{N}\cos\psi d\psi
  \label{u}
\end{equation}
with constants
\begin{equation}
    \frac{1}{L} =
\frac{1}{2}\left(\frac{1}{r_{+}}+\frac{1}{r_{-}}\right)
 = \frac{1}{a(1-e^2)}, \qquad \frac{1}{N} =\frac{1}{2} 
    \left(\frac{1}{r_{+}}-\frac{1}{r_{-}}\right)
     = \frac{-e}{a(1-e^2)}
\label{LA}
\end{equation}
yields $(u_- -u)(u-u_+)=N^{-2}\cos^2 \psi$, so the relation \eqref{diffpsi2} is immediately integrated giving the general result applicable for computation of precession in any isotropic metric:
\begin{equation}
    \phi(r) - \phi(r_{-}) = \frac{J}{\sqrt{r_{-}r_{+} \left(
    E-1\right)}}\int_{\psi_{-}}^\psi d\psi =
   \frac{J\sqrt{u_{-}u_{+}}  }{\sqrt{ \left(
    E-1\right)}}(\psi
    -\psi_{-}) = C^{-\frac{1}{2}}\left(\psi(r) - \psi_{-}\right),
    \label{g}
\end{equation}
where the parameter $C$, in accord with \eqref{C} and \eqref{EJ}, is
treated as a function of $r_+$ and $r_{-}$. It is convenient to express $\psi$ in terms of $u$ or $r$ (see \eqref{u}, \eqref{LA}), i.e.
\begin{equation}
\psi(r)=\arcsin\frac{u(r)-\frac{1}{2}(u_{+}+u_{-})}{\frac{1}{2}(u_{+}-u_{-})}
= \arcsin\frac{a \left(1-e^2\right)-r}{-e r} = \arcsin\left( \frac{N}{r} - \frac{N}{L}\right).
\label{psir}
\end{equation}
So, \eqref{g} might be presented directly as a function of $r$:
\begin{equation}
\phi(r) - \phi(r_{-}) = C^{-\frac{1}{2}}\left[\arcsin\left( \frac{N}{r} - \frac{N}{L}\right) + \frac{\pi}{2}\right],
\label{G}
\end{equation}
with $\psi_{-} \equiv \psi(r_{-})=\arcsin(-1) =
-\frac{\pi}{2}$ at perihelion, and $\psi_{+}=\frac{\pi}{2}$ at aphelion . 

In particular, the angular shift of the perihelion point (precession) per one revolution (from $r_{-}$ to $r_{+}$ and back) is:
\begin{equation}
    \Delta \phi = 2 \left|\phi(r_{+}) - \phi(r_{-})\right| - 2\pi = 
    2 C^{-\frac{1}{2}}\left|\psi_{+}-\psi_{-}\right| -2\pi = 2\pi
    \left(C^{-\frac{1}{2}}-1\right),
\label{precessfinal}
\end{equation}
where the expression for $C^{-1/2}$ follows from \eqref{C}:
\begin{equation}
    C^{-\frac{1}{2}} = \left(\frac{
            B_{-}-B_{+}}{(1-B_{-})
            \frac{B_{+}}{ D_{+}}
                        \frac{ r_{-}  }{ r_{+} }- (1-B_{+})
                        \frac{B_{-}}{ D_{-}}
                        \frac{ r_{+} }{
                r_{-} }}\right)^{\frac{1}{2}}.
\label{Cgeneral}
\end{equation}

\section{Distinction between vacuum and scalar background}\label{sec4}

To distinguish the precession effect in scalar background generated by the Papapetrou
space-time from that in the Schwarzschild vacuum we should apply in final
expressions the expansion of metric coefficients up to the second order in $M/r$:
\begin{equation}
 D_P(r) = B_P^{-1}(r) = e^{2M/r} \simeq 1 + \frac{2M}{r} + \frac{2M^2}{r^2}  + ... \quad  
 \label{series1}
\end{equation}
in scalar background \eqref{P}, and
\begin{equation}
 D_S(r)= \left(1+\frac{M}{2 r} \right)^4\simeq 1 + \frac{2M}{r} +
 \frac{3M^2}{2r^2} + ...\quad ,
\end{equation}
\begin{equation}
 B_S^{-1}(r) = \left( \frac{ 1-\frac{M}{2 r}}{1+\frac{M}{2 r}}
 \right)^{-2} \simeq 1 + \frac{2M}{r} + \frac{2M^2}{r^2} + ...   
 \label{series4}
\end{equation}
in vacuum \eqref{S}. Substitution of these into
\eqref{precessfinal}-\eqref{Cgeneral} shows that in both
cases the expressions for the perihelion precession
per revolution coincide to the first order:
\be \label{phipredicted}  \Delta \phi_{P} = \Delta \phi_S = \frac{6\pi M }{ L} + ...  ,  \ee 
where $L$ \eqref{LA} is the semi-latus rectum for the orbit in
corresponding isotropic coordinates. The difference between vacuum and
scalar cases is revealed only in the second order of $M/r$.
Expanding $C^{-1/2}$ \eqref{Cgeneral} into power series with respect to $u_{-}$ and $u_{+}$,
and expressing the latter via $L$ and $N$ from \eqref{LA},
\begin{equation}
    u_{-} =  \frac{N-L}{N L}, \qquad u_{+} = \frac{N+L}{N L},
\end{equation}
we get, after some algebra:
\begin{equation}
    \Delta\phi_S = 2\pi \left(C^{-1/2}_S-1\right) = \frac{6 \pi  M}{L}
    +\frac{25 \pi  M^2}{2 L^2} -\frac{9 \pi  M^2}{2 N^2}+ \mathcal{O}\left(\frac{M^3}{r^3}\right)
    \label{1} 
\end{equation}
and
\begin{equation}
    \Delta\phi_P = 2\pi \left(C^{-1/2}_P-1\right) = \frac{6 \pi
    M}{L}+\frac{41 \pi  M^2}{3 L^2}  -\frac{4 \pi  M^2}{N^2} +
    \mathcal{O}\left(\frac{M^3}{r^3}\right).
    \label{2}
\end{equation}
Thus, the angular shift per revolution is larger in case of scalar background by
\begin{equation}
    \Delta\phi_P - \Delta\phi_S= \frac{1}{6} \pi  M^2
    \left(\frac{7}{L^2}+\frac{3}{N^2}\right).
    \label{3}
\end{equation}

So, we have provided second-order corrections \eqref{1} and \eqref{2} to the standard
first-order value \eqref{phipredicted}, and the difference between vacuum and
scalar cases \eqref{3}.

\section{Juxtaposition with the curvature coordinates}\label{sec6}

It is interesting to compare our results with those obtained in traditional curvature
coordinates for the Schwarzschild solution,
 \be  \label{7a}  ds^2= b(R) dt^2 - a(R) dR^2 - R^2 \left( d\theta^2
 +  \sin^2 \theta d\phi^2\right) \ee
 with
\be  b(R)= \Big(1- \frac{2 M}{R}  \Big),  \qquad a(R)= \Big(1- \frac{2 M}{R}  \Big)^{-1} ,\ee 
(see, e.g.,  \cite{Weinberg1972}). In fact, isotropic relations obtained above might also be
found from these results by direct
substitution of the transformation from curvature to isotropic coordinates:
\begin{equation}
R= r\left(1+\frac{M}{2r}\right)^2 \, \leftrightarrow \, R^2 = r^2 D_S(r).
\label{trans}
\end{equation}
In particular, the known solution for the curvature coordinates
 \be \label{20}   \phi=   \pm \int  \frac{\sqrt{a} \,\,  dR}{ R^2  \Big[   1/(J^2 b) - E/J^2- 1/R^2    \Big]^{1/2}  }\,  \ee
transforms via \eqref{trans} into \eqref{19}.
The constants \eqref{EJ} can also be obtained by the same procedure from their curvature analogs:
\be   \label{EJS}    E= \frac{ \frac{R_{+}^2  }{ b^{+} }   -
	\frac{R_{-}^2  }{ b^{-} }    }{R_{+}^2 - R_{-}^2  },  \qquad  J^2 =  \frac{  \frac{1}{b^{+}}- \frac{1}{b^{-}}  }{
	\frac{1}{R_{+}^2  }   -  \frac{1}{R_{-}^2  } }, \,\,\, \text{etc}.  \ee
All this might serve as independent verification of our calculations.

The discrepancies between the two approaches are observed only after integration is involved. Indeed, the general integral result for the precession effect per arbitrary angular position in curvature coordinates
\cite{Weinberg1972},
\begin{equation}
\phi(R)-\phi(R_{-}) = \left( 1 + 3 \frac{M}{L}  \right) \left(
\psi - \psi_{-} \right)- \frac{M}{L}\left( \cos\psi -
\cos\psi_{-} \right),
\label{l}
\end{equation}
proves to be functionally distinct (non-linear in $\psi$) even in the
first order from the analogous expression \eqref{g} in isotropic
coordinates which is linear in $\psi$ both in vacuum and scalar
background, without periodic trigonometric terms. These
terms might be responsible for additional acceleration or deceleration of
a test body in different parts of the orbit if measured in curvature
coordinates (only for integer number of revolutions they do not impact the final formulas for the precession of perihelion).

It stands to reason that the adequate choice of coordinates matters.
In this respect, it is worth to note that Fisher in his seminal work \cite{Fisher1948} looked for the scalar solution namely in curvature coordinates, but his results (i.e. coefficients of metric and scalar potential) proved to be defined only up to conformity to some non-trivial algebraic condition.

 On the other hand, transfer from isotropic Papapetrou
metric \eqref{P} to corresponding (antiscalar) curvature coordinates contains the Lambert function $W(x)$: 
\begin{equation}
R=R(r)=re^{M/r} \,\,\,\,\Rightarrow \,\,\,\,
r= r(R) = -M/W\left(-\frac{M}{R}\right), 
\label{invers}
\end{equation}
and yields specific (due to non-unique behaviour of the $W$-function) interval
\begin{equation}
	ds^2 = e^{2W\left(-\frac{M}{R}\right)}dt^2 - \left( 1 + W\left( -\frac{M}{R}\right)\right)^{-2}dR^2 - R^2 d\Omega^2,
	\label{PapaLambert}
\end{equation}
with simultaneous transfer from Newtonian potential to its curvature
counterpart \cite{mm20},
\begin{equation}
\varphi(R) = -W(-M/R),
\label{phiLambert}
\end{equation}
being of rather obscure physical meaning.

\section{Precession of perihelia}\label{sec7}

As a first example we consider Mercury. The values of perihelion and aphelion in this case are $r_{-}=46001200$~km
and  $r_{+}=69816900$~km, hence $L \simeq 5.55 \times 10 ^{7} $~km and
$N \simeq -26.97 \times 10^{7}$~km. There are $\sim 415$ revolutions
per century in this case, so the value of $ \Delta \phi_{GR}$
predicted by \eqref{phipredicted} (which leads to $42.96$ arcsec per
century vs. observed $\Delta \phi_{obs} = 43.11 \pm 0.21 $
arcsec ) gains the second-order contribution of about $10^{-7}$ arcsec
per century, which is, sure enough, observationally negligible. With
that, the difference \eqref{3} between scalar and vacuum backgrounds
for the case of Mercury amounts to $\sim 2.26\times 10^{-7}$ arcsec per
century.

While for the Mercury the contribution of the second-order terms into
expected observational effects is negligible, the situation might be different in stronger fields, e.g., for stars moving
close to supermassive compact objects in the centers of galaxies
\cite{abuter_detection_2020}, and also for orbits of test bodies with
large eccentricities.  In this respect we now apply our approach to the perihelion precession phenomenon of four stars in the S-cluster in the center of the Milky Way -- S2, S38, S55, S62 (see \cite{2017ApJ...837...30G}, \cite{2020ApJ...889...61P}).

 \begin{table*}[ht]
 	\caption{Orbital parameters and perihelion shift for Mercury and four S-stars. Numbers in columns (1 rev) and (100 year) imply values per one revolution and per one century, respectively.}
 	\label{tabular: tab1}
 	\begin{center}
            \setlength{\tabcolsep}{.05cm}
 		\begin{tabular}{cccccccccc}
 			\hline
 			\hline
 			\makecell{Object} &  \makecell{$a$,\\ AU} &  $e$ & \makecell{$T$, \\years } & \multicolumn{2}{c}{$\Delta \phi_S$} & \multicolumn{2}{c}{$\Delta \phi_P$} & \multicolumn{2}{c}{$\Delta \phi_P - \Delta \phi_S$}\\
 			\hline
            {} & {} & {} & {} & (1 rev) & (100 yrs.) & (1 rev) & (100 yrs.) & (1 rev) & (100 yrs.)\\ [5pt]
                \hline

                Mercury & 0.3871 & 0.2056 & 0.24 & 0.1033$''$ & 43.0541$''$ & 0.1033$''$ & 43.0531$''$ & 5.43$\times10^{-10}$$''$ & 2.26$\times10^{-7}$$''$\\ [5pt]
 			
 			  S2 & 970 & 0.8839 & 16.00 & 12.6431$'$ & 79.0195$'$ & 12.6438$'$ & 79.0235$'$ & 0.0384$''$ & 0.2340$''$\\ [5pt]
 			
 			 S38 & 1022 & 0.8201 & 19.20 & 8.0149$'$ & 41.7441$'$ & 8.0151$'$ & 41.7454$'$ & 0.0149$''$ & 0.0776$''$\\ [5pt]
 			
 		 	 S55 & 780 & 0.7209 & 12.80 & 7.1591$'$ & 55.9307$'$ & 7.1593$'$ & 55.9322$'$ & 0.0113$''$ & 0.0881$''$\\ [5pt]
 			
 			S62 & 740 & 0.9760 & 9.90 & 76.5349$'$ & 773.0800$'$ & 76.5596$'$ & 773.3290$'$ & 1.4804$''$ & 14.9532$''$\\ [5pt]
 			\hline
 				\end{tabular}
 			\end{center}
 	\end{table*}

Table ~\ref{tabular: tab1} presents the orbital parameters of
specified objects and values of perihelion shift. Among the selected objects, the perihelion
shift effect is most pronounced for S62 due to its remarkable
eccentricity and close proximity to the central object (short period). For
the same reasons, the difference in values $\Delta \phi_P - \Delta
\phi_S $ for S62 is essentially greater in comparison to
other cases. 

Regarding the S2 star, we can estimate the significance of distinction
between vacuum and scalar background based on observational data. To
be more precise, the precession of S2 orbit was detected after 27
years monitoring of its motion \cite{abuter_detection_2020}. Authors
inferred that orbit precession corresponds to $\Delta \phi = f_{SP}
\times \delta \phi$ per one revolution, where $\delta \phi = 12.1'$ is a
robustly detected value, $f_{SP} = 1.10 \pm 0.19$ is a dimensionless
parameter determined through posterior fitting procedure and
detailed analysis. Therefore, perihelion shift for S2 amounts to
$\Delta \phi = 13.31' \pm 2.30'$, yielding an observational
measurement error of $4.60'$, which is several orders of magnitude
larger than the obtained difference $\Delta \phi_P - \Delta \phi_S =
0.03841''$. This fact indicates that for the given star the current
accuracy of observational instruments is insufficient to detect the
distinction between vacuum and scalar background in
weak-field regime.

Situation might be more optimistic, for example, in
the case of S62 star, as for it the difference $\Delta \phi_P - \Delta
\phi_S $ is almost 2 orders of magnitude larger than for S2 star.

The results obtained analytically in the second approximation can be verified by direct numerical integration in Eq.~\eqref{19} with exact metric coefficients \eqref{P}-\eqref{S}, which we have performed using Wolfram Mathematica. Then, in the case of Mercury integration yields almost the same but slightly smaller effects than those predicted with analytic second approximation, as seen in Table ~\ref{tabular: tab2}.

\begin{table*}[ht]
 	\caption{The values of perihelion shift obtained through numerical integration. }
 	\label{tabular: tab2}
 	\begin{center}
            \setlength{\tabcolsep}{.05cm}
 		\begin{tabular}{ccccccc}
 			\hline
 			\hline
 			\makecell{Object} & \multicolumn{2}{c}{$\Delta \phi_S$} & \multicolumn{2}{c}{$\Delta \phi_P$} & \multicolumn{2}{c}{$|\Delta \phi_P - \Delta \phi_S|$}\\
 			\hline
            {} & (1 rev.) & (100 yrs.) & (1 rev.) & (100 yrs.) & (1 rev.) & (100 yrs.)\\ [5pt]
                \hline

                Mercury & 0.1033$''$ & 43.0405$''$ & 0.1033$''$ & 43.0405$''$ &  9.25$\times10^{-10}$$''$ & 3.85$\times10^{-7}$$''$  \\ [5pt]

                S2 & 12.6466$'$ & 79.0413$'$ & 12.6475$'$ & 79.0474$'$ & 0.0590$''$ & 0.3687$''$  \\ [5pt]

                S38 & 8.0163$'$ & 41.7514$'$ & 8.0167$'$ & 41.7534$'$ & 0.0232$''$ & 0.1207$''$  \\ [5pt]

                S55 & 7.1602$'$ & 55.9394$'$ & 7.1605$'$ & 55.9418$'$ & 0.0179$''$ & 0.1397$''$  \\ [5pt]

                S62 & 76.6637$'$ & 774.3805$'$ & 76.7011$'$ & 774.7584$'$ & 2.2446$''$ & 22.6733$''$  \\ [5pt]
 			\hline
 				\end{tabular}
 			\end{center}
 	\end{table*}

The same calculations for the S-cluster stars with appreciably higher values of eccentricity $e$ yield almost the same but, on the contrary, somewhat larger values of effects than those in the second approximation approach. On the whole, the largest difference in the effect between scalar and vacuum background is found for the most high-eccentric object S62, but still the value of $\sim22$ arcsec per century is hardly approachable observationally.

\section{Shadow effect}\label{sec8}

In this section we reconsider the calculation of the strong-field shadow effect in vacuum and scalar background which we performed earlier in Ref. \cite{mm18} based on the complexification of the potential resulting in transfer from the JNW metric to antiscalar regime. Here we proceed from the assumption that the underlying isotropic metric \eqref{is} can be deduced without any complexification, making the results more physically tractable.

In general, the static spherically symmetric interval can be expressed as
\begin{equation}
	ds^2 = g_{\alpha \beta} dx^{\alpha} dx^{\beta} = g_{tt}(r)dt^2 + g_{rr}(r)dr^2 + g_{\theta \theta}(r)d\theta^2 + g_{\phi \phi}(r,\theta)d\phi^2,
	\label{GeneralMetric}
\end{equation}
where $g_{\phi \phi}(r,\theta)=g_{\theta \theta}\sin^2{\theta}$, and
the signature (+~\textendash~\textendash~\textendash) is adopted. In the context of the shadow effect, one starts with considering the deflection of light. The deflection angle for the nearest approach
distance $r_0$ is given by \cite{virbhadra1998role}:
\begin{equation}
	\hat{\alpha}(r_0) = \int\limits_{r_0}^\infty\left(\frac{g_{rr}(r)}{g_{\theta \theta}(r)}\right)^{1/2}\left(\frac{g_{\theta \theta}(r)}{g_{\theta \theta}(r_0)} \frac{g_{tt}(r_0)}{g_{tt}(r)} - 1\right)^{-1/2} dr - \pi.
	\label{DeflectionAngle}
\end{equation}

Another important quantity is the impact parameter $\mathcal{J}$ which is connected with the distance from the center of the lens to the observer $D_l$ through the relation $\mathcal{J}=D_l \sin{\Theta}$, where $\Theta$ is the observer's polar angle for image of the source. It can be written in terms of the general metric \eqref{GeneralMetric} in the following form:
\begin{equation}
	\mathcal{J}=\mathcal{J}(r_0)=\left(\frac{-g_{\theta \theta}(r_0)}{g_{tt}(r_0)}\right)^{1/2}.
	\label{ImpactParameter}
\end{equation}

Now, calculation of impact parameter for \eqref{is} with \eqref{P} and \eqref{S} yields 
\begin{equation}
	\mathcal{J}_P = r_0 \exp{\left(\frac{2M}{r_0}\right)} = 1+\frac{2M}{r_0}+\frac{2M^2}{r_0^2}+\mathcal{O}\left(\frac{M^3}{r_0^3}\right)
	\label{ImpactParameterScalar}
\end{equation}
and
\begin{equation}
	\mathcal{J}_S = \frac{r_0 \left(1+\frac{M}{2r_0}\right)^3}{1-\frac{M}{2r_0}} =  1+\frac{2M}{r_0}+\frac{7M^2}{4r_0^2}+\mathcal{O}\left(\frac{M^3}{r_0^3}\right),
\label{ImpactParameterVacuum}
\end{equation}
correspondingly, which coincide only at first order as typically should be for any general-relativistic  effects in strong fields. 

In region where the impact parameter reaches its critical value, a photon sphere emerges. The existence of the photon sphere is related to the maximization of the deflection angle. As a light ray approaches the photon sphere, its trajectory experiences bending, when the derivative of deflection angle \eqref{DeflectionAngle} with respect to $r_0$ is zero, i.e.
\begin{equation}
	g_{tt}\frac{\partial g_{\theta \theta}}{\partial r_0} = g_{\theta \theta}\frac{\partial g_{tt}}{\partial r_0}.
	\label{EqPhotonSphere}
\end{equation}
Alternatively, Eq.~\eqref{EqPhotonSphere} might be obtained using the condition
\begin{equation}
	\frac{\partial}{\partial r_0}\mathcal{J}^2(r_0)=0.
	\label{CritImpCondition}
\end{equation}
Solving \eqref{EqPhotonSphere} provides the exact value for the radius of the photon sphere denoted as $r_0 = r_{ps}$. Then for scalar background one obtains:
\begin{equation}
	r_{ps}^P = 2M,
	\label{PSScalar}
\end{equation}
while for the Schwarzchild isotropic metric we get four solutions:
\begin{equation}
	r_{ps}^S = \pm \frac{M}{2}, \quad r_{ps}^S = \left(1 \pm \frac{\sqrt{3}}{2}\right)M,
	\label{PSVAcuum}
\end{equation}
where only the maximal positive value does not meet any contradictions.

The shadow radius $R_{sh}$ is determined by evaluating the impact parameter at the photon sphere radius $\mathcal{J}(r_{ps})$. In this way, for vacuum in isotropic coordinates, it proves to be $R_{sh}^S = 3\sqrt{3}M = 5.196M$. At the same time, for scalar background we get the value $R_{sh}^P = 2eM = 5.437M$ which is 5\% larger than in vacuum. 

This 5\% difference is due to appreciable distinction between the second order terms in the decompositions \eqref{ImpactParameterScalar} and \eqref{ImpactParameterVacuum} with respect to $M/r_0$. Thus, to draw a final conclusion here one needs the value of the mass of the central compact object $M$ measured by independent methods (e.g., via the orbits of surrounding test objects) with the accuracy significantly higher than the one currently available.

\section{Conclusion}\label{sec9}

Using isotropic coordinates \eqref{is}, we have calculated the perihelion shift in antiscalar background for the Mercury and four S-cluster stars in the vicinity of Sgr A*, and compared the results to the case of Schwarzschild vacuum background. The distinction is most pronounced for the S62 star, but even in this case with the accumulated difference of $\sim 22$ arcsec per century, it is, in practice, observationally imperceptible. At the same time, in stronger fields, we have assessed the shadow size of the compact objects like the one in the center of the Milky Way, obtaining much more appreciable difference of $\sim 5\%$. However, in this case, to draw the conclusion, one has to know the value of the central object's mass obtained with an independent method with the accuracy within a percent at most.

Hence, both possibilities are currently viable from an observational viewpoint, leaving open the question of the nature of the vacuum we live in. If observations will ultimately favor antiscalar background, the minimal admissible Einstein equations should be upgraded from $R_{\mu\nu}=0$ to $R_{\mu\nu}=2\epsilon \varkappa \varphi_{,\mu} \varphi_{,\nu}$, with $\epsilon=-1$, with the ensuing reinterpretation of gravity within general relativity.

\section*{Acknowledgements}
This research is funded by the Science Committee of the Ministry of Science and Higher Education of the Republic of Kazakhstan (Grant No. AP19678165 and Program No. BR21881880). The authors thank Taras Panamarev for some useful ideas.

\bibliography{sn-bibliography}

\end{document}